\documentstyle{ichep}
\input epsf.sty
\def\lsim{\mathrel{\raise.2ex\hbox{$<$}\hskip-.8em\lower.9ex\hbox{$\sim$}}}
\def\gsim{\mathrel{\raise.2ex\hbox{$>$}\hskip-.8em\lower.9ex\hbox{$\sim$}}}

\begin{document}

\title{Supersymmetric Unification: a mini-review of recent developments$^*$}

\author{V. Barger$^{\dag }$, M. S. Berger$^{\S }$,
and P.~Ohmann$^{\dag }$}

\affil{\dag\ Physics Department,
University of Wisconsin, Madison, Wisconsin 53706\\
\S\ Physics Department, Indiana University, Bloomington, Indiana 47405}

\abstract{\centering
Some recent results in supersymmetric grand unified theories are
reviewed.}

\twocolumn[\maketitle]

\fnm{7}{Invited talk given by V. Barger at the 27th International Conference on
High Energy Physics, Glasgow, 20--27th July 1994.}

\section{Introduction}
The Standard Model (SM) of particle interactions has proven very successful in
describing collider physics. To go beyond,  one is interested in discovering
new states or eliminating
free parameters of the SM. Usually each of these avenues
involves  new or larger symmetries. The idea of
supersymmetric (SUSY) grand unified theories (GUTs) is no exception:
the gauge group of the Standard Model is included in a larger grand unified
group and a new symmetry relating particles of different spin
is introduced. This approach then predicts the existence of many new
states (the sparticles) and can eliminate free parameters (e.g.
$\sin \theta _W$) that exist in the Standard Model.

Not long after the unification of the electroweak interactions
the idea of a further unification of all the forces in the Standard Model
into a single gauge group was born\cite{gg}. Shortly thereafter the
implications for the value of the weak mixing angle was
derived\cite{gqw}. The low-energy theory was then made supersymmetric\cite{drw}
yielding a slightly different value for $\sin \theta _W$. Before the precision
data from LEP both these versions of a grand unified theory incorporating
a desert between the electroweak and the GUT scale were experimentally
viable. With the recent precise electroweak data,  the
supersymmetric version alone can describe the experimental results in the
absence of an intermediate scale\cite{amaldi}.


The other development that is central to the success of
SUSY unification is electroweak symmetry breaking.
Supersymmetry protects the hierarchy of the electroweak scale and the GUT
scale; in addition the breakdown of the electroweak symmetry can be understood
when  the supersymmetry is local\cite{nath}. The large
value of the top quark Yukawa coupling enhances large logs that
cause the spontaneous breakdown of the electroweak
symmetry\cite{ir}.

Following the revival of
interest in SUSY GUTs caused by  the unification of the gauge couplings,
some  of the more model-dependent earlier predictions  were reinvestigated. In
particular, the relation between the bottom quark
and the tau lepton mass that occurs in some minimal versions of grand unified
theories was updated. The equality of $b$ and $\tau$ Yukawa couplings at the
GUT scale was first proposed in Ref.~\cite{ceg}. The importance of a large
top Yukawa coupling in suppressing the bottom quark mass was emphasized by
Iba\~{n}ez and Lopez\cite{ibanez}. More recently the correlation between
the bottom quark mass and the tau lepton mass was explored, and an
inconsistency in the nonsupersymmetric desert prediction was
uncovered\cite{hgs}.  Moreover $b$-$\tau$ coupling unification is perfectly
viable in the minimal supersymmetric extension to the Standard Model. While
Yukawa coupling unification is
not as general as that of gauge coupling unification (it involves specific
assumptions about the GUT symmetry breakdown), the success of the simplest
relation added extra impetus to the
interest in supersymmetric GUTs.
The $m_b/m_{\tau }$ relation  implies that the top quark Yukawa coupling is
probably
at its infrared quasi-fixed point, which in turn limits the relation of the
top quark and the ratio of the Higgs vevs in supersymmetry to a narrow corridor
of values\cite{bbo}--\cite{lp}. When the uncertainty on the top quark mass from
CDF is reduced, then only small ($\approx 1$) and
large ($\approx m_t/m_b$) values of $\tan \beta $ will be allowed, with the
small $\tan\beta$ solution preferred.
[The situation is complicated somewhat by the presence of threshold corrections
at the electroweak and grand unified scales---if large enough and if
$\alpha _3(M_Z^{})$ is at the lower end of its experimentally allowed range,
then the
fixed-point solution might not apply.] The relationship implied
by the top quark Yukawa coupling fixed point on the value of $\tan \beta$ also
occurs in a top quark condensate mode\cite{bccsw} when the scale $\Lambda $ is
near the GUT scale.

\section{Evolution of Dimensionless Couplings}

The gauge couplings evolve according to renormalization group equations (RGE)
with the solution
\begin{equation}
\alpha _i^{-1}(Q)=\alpha _i^{-1}(M_G^{})-{{b_i}\over {2\pi}}t \;.
\end{equation}
at the one-loop level where $t=\ln (Q/M_G)$ defines the scale.
The parameters $b_i$ are determined by the particle content of the
effective theory.

The evolution of the top quark Yukawa coupling is described by the RGE,
\begin{equation}
{{d\lambda _t}\over {dt}}={{\lambda _t}\over {16\pi^2}}\left [
-{13\over 15}g_1^2-3g_2^2-{16\over 3}g_3^2+6\lambda _t^2+\lambda _b^2\right ]
\;.
\end{equation}
Figures 1 and 2 show
the solution
of these renormalization group equations for values of the bottom quark running
mass. One sees that the top Yukawa coupling is driven to its
infrared fixed point\cite{bbo}--\cite{lp3}.

\begin{center}
\epsfxsize=2.5in
\hspace{0in}
\epsffile{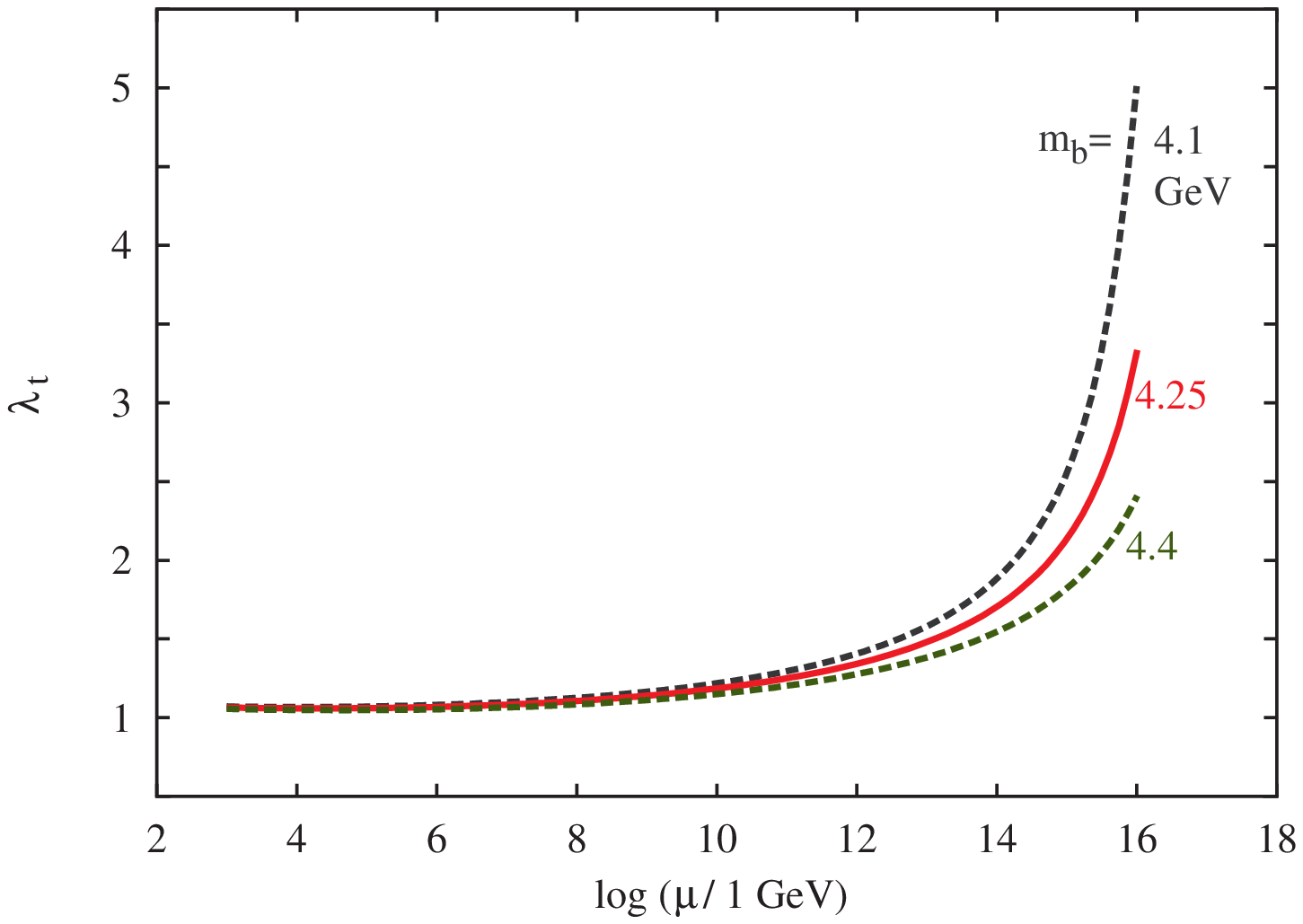}

\smallskip
\parbox{3.25in}{\small Fig.~1. If $\lambda _t$ is large at $M_G^{}$, then
the renormalization group equation causes $\lambda _t(Q)$ to evolve rapidly
towards an infrared fixed point as $Q \rightarrow m_t$ (from Ref.~\cite{bbo}).}
\end{center}

\smallskip

\begin{center}
\epsfxsize=2.5in
\hspace*{0in}
\epsffile{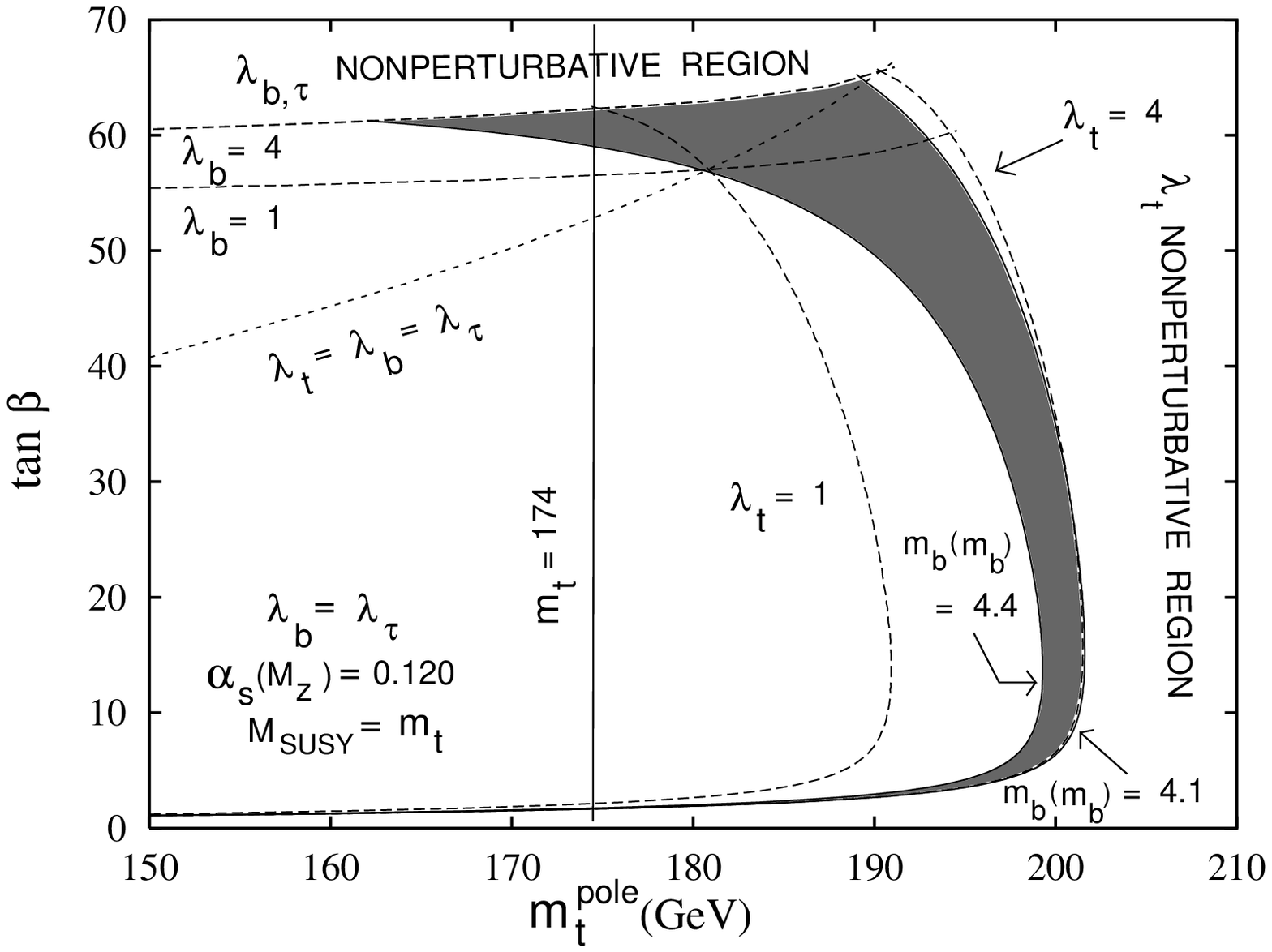}

\smallskip
\parbox{3.25in}{\small Fig.~2. Contours of constant $m_b(m_b)$ in the
$m_t(m_t),\tan\beta$ plane with contours of constant GUT scale Yukawa
couplings (adapted from Ref.~\cite{bbo}).}
\end{center}

Converting this relation for the top quark pole mass yields\cite{bbop}
\begin{equation}
m_t^{\rm pole }\approx(200 {\rm GeV})\sin \beta \;.
\end{equation}
The value of 200 GeV that appears in the above equation is subject to
some uncertainty due to varying $\alpha _3(M_Z^{})$ and to threshold
effects\cite{lp,thr}.

Adding a singlet $N$ to the minimal supersymmetric model
provides an additional coupling $\lambda H_1H_2N$ in the superpotential,
which could conceivably suppress the bottom quark mass sufficiently that the
top quark
Yukawa coupling could be reduced. However constraints on the perturbativity of
the coupling $\lambda $ preserves the fixed point condition\cite{nmssm,lp3}.
A four generation model is not easily compatible with Yukawa
unification\cite{gmp}.

\section{RGE Evolution of Sparticle Masses}

An attractive property of models based on supergravity is that the
symmetry breakdown in the electroweak sector can be attributed to large
logs that contribute to the Higgs potential\cite{grz}--\cite{arnowittnath}.
One must arrive at the correct
scale for the electroweak interactions without breaking color or charge. This
is accomplished by imposing two
minimization conditions obtained from the Higgs potential.

The minimum of the Higgs potential must occur by the acquisition
of vacuum expectation values.
Minimizing the tree-level potential with respect to the two neutral CP-even
Higgs degrees of
freedom yields the two conditions
\begin{eqnarray}
{1\over 2}M_Z^2&=&{{m_{H_1}^2-m_{H_2}^2\tan ^2\beta }
\over {\tan ^2\beta -1}}-\mu ^2 \;.  \label{treemin1} \\
-B\mu &=&{1\over 2}(m_{H_1}^2+m_{H_2}^2+2\mu ^2)\sin 2\beta \;.
\label{treemin2}
\end{eqnarray}
where $m_{H_1}^{}$ and  $m_{H_2}^{}$
are soft-supersymmetry breaking parameters and $\mu $ is the Higgs mass in
the superpotential.
In order to avoid large cancellations between the terms on the right-hand side
of Eq.~(\ref{treemin1}), some naturalness criteria are inposed, which in turn
typically imply that the sparticle masses are not too high ($\lsim1$~TeV).

For the low $\tan\beta$ fixed-point solution as few as two inputs for the
soft-supersymmetry breaking parameters are needed---a universal scalar mass
$m_0$ and a common gaugino mass $m_{1/2}$\cite{copw}. The low energy sparticle
mass can be given in terms of these inputs.

The heaviest chargino and the two heaviest neutralino states are primarily
Higgsino with masses approximately equal to $|\mu |$.
Typical mass relationships are displayed in Figure 3.

\begin{center}
\epsfxsize=2.5in
\hspace*{0in}
\epsffile{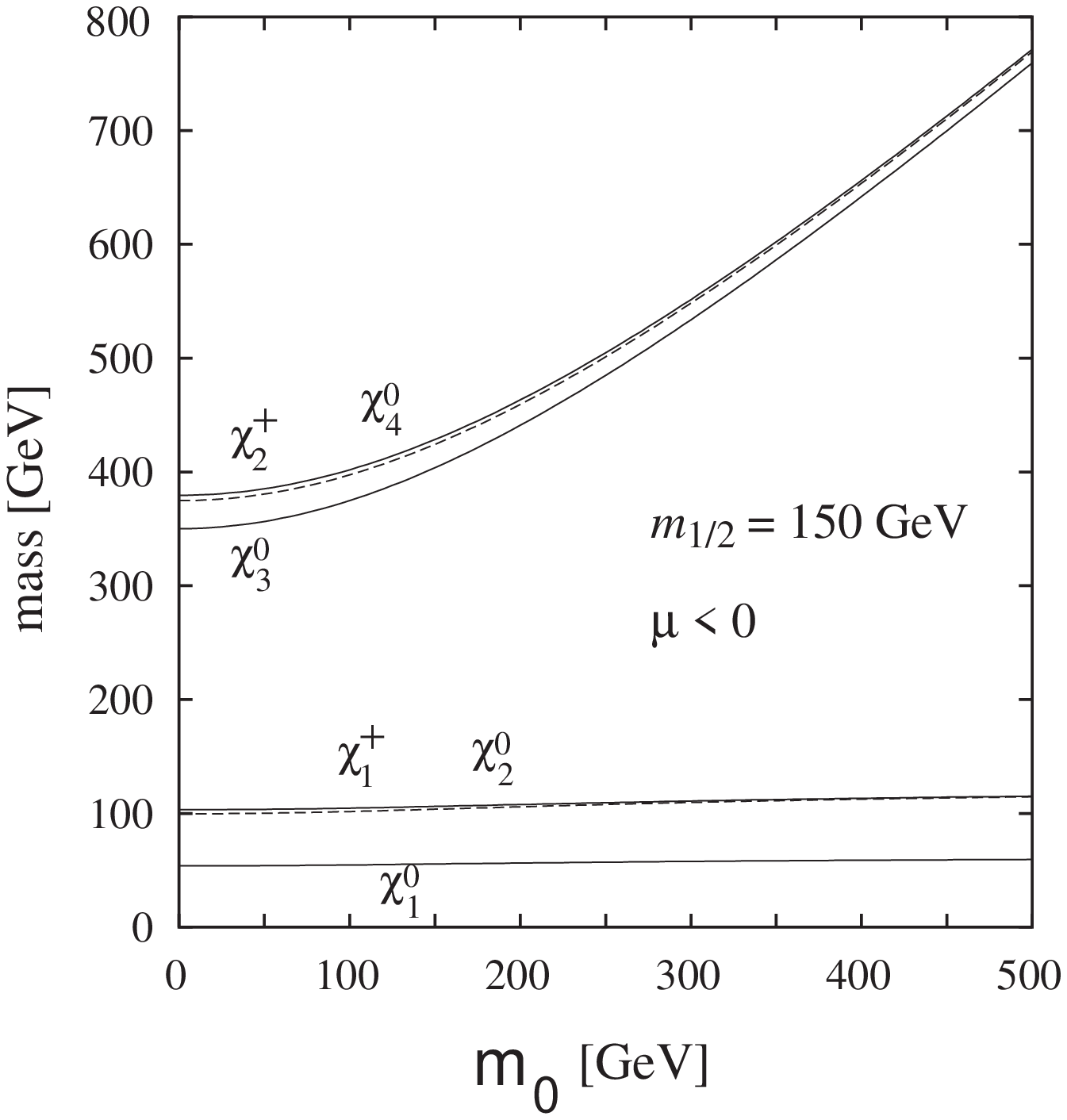}

\parbox{3.25in}{\small Fig.~3. The chargino and neutralino masses are plotted
versus $m_0^{}$ for $m_{1/2}=150$ GeV for a low value of $\tan \beta$ and
$\mu < 0$ (from Ref.\cite{madph801}). }
\end{center}

For the low-$\tan \beta $ solution, the mass of the lightest Higgs $h$
comes mainly from radiative corrections\cite{bbop,lp3,dh,hh,ceqr,lnpwz}.
Experiments at LEP\,II will cover the region $m_h\lsim\sqrt s/2$.
Recent work\cite{msw} has shown that a $h\to b\overline{b}$ search at the
Tevatron may be possible. The heavy Higgs states are
(approximately) degenerate $\approx M_A$; see Figure~4.
The Higgs discovery potential at $e^+e^-$ colliders has recently been
discussed\cite{andre}.

The squark and slepton masses also display
simple asymptotic behavior at large $|\mu|$; see Figure 5.
The first and second squark generations are approximately degenerate.
The splitting of the stop quark masses grows as $|\mu|$ increases and the
lightest stop can be as light as 45 GeV (or even lighter with fine tuning).
The masses could be much larger
than is indicated in the figure since the value of $m_0$ could be large.

While the universality of the scalar masses has been assumed for the Figures
presented above, recently there has been much interest in considering
the implications of nonuniversality on the supersymmetric spectrum and on
reconsidering the constraints from flavor changing neutral
currents\cite{susy94}.

\begin{center}
\epsfxsize=2.5in
\hspace*{0in}
\epsffile{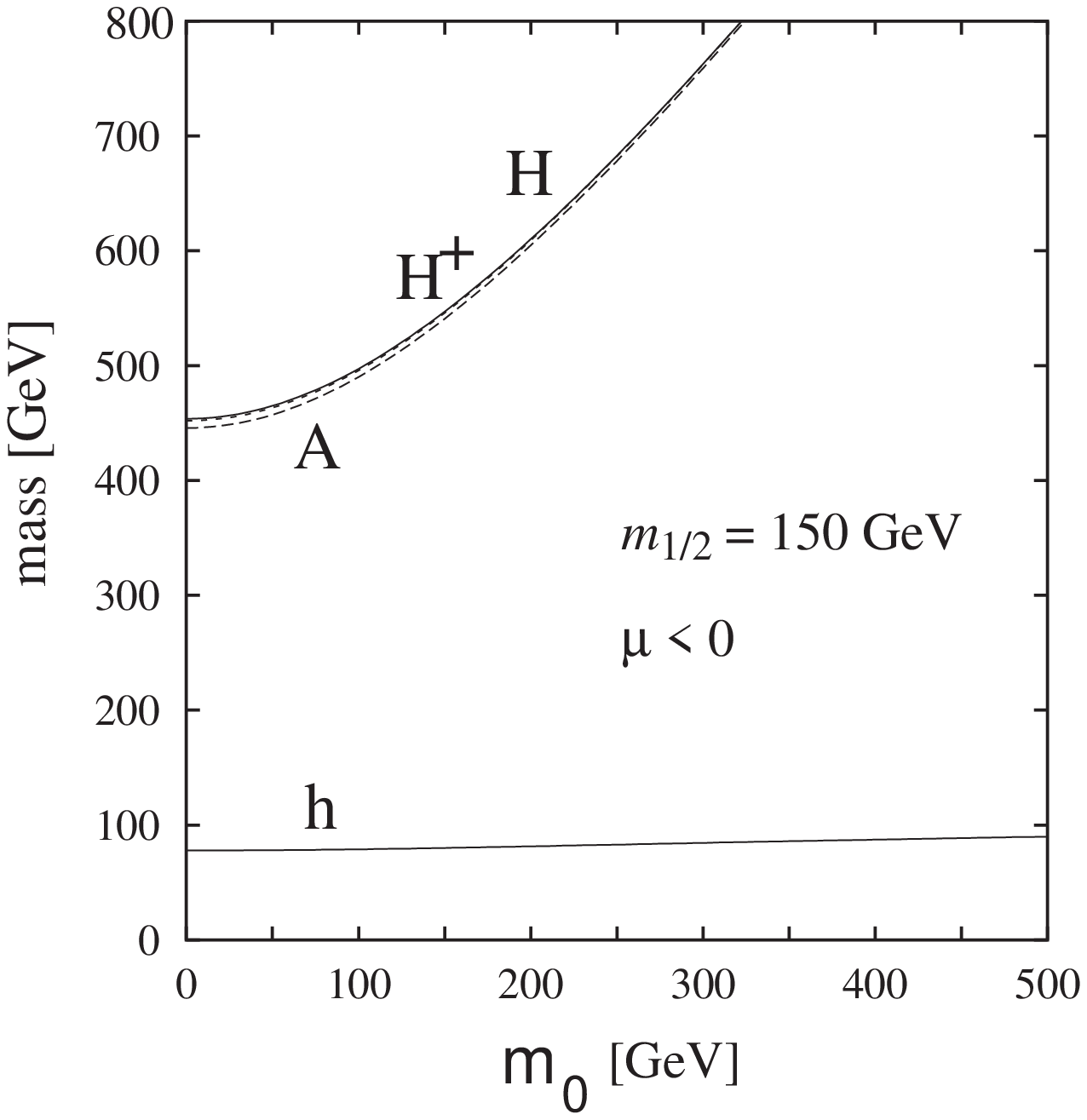}

\parbox{3.25in}{\small Fig.~4. The supersymmetric Higgs masses are plotted
versus $m_0^{}$ for $m_{1/2}=150$ GeV for a low value of $\tan \beta$ and
$\mu < 0$ (from Ref.\cite{madph801}).}
\end{center}

\begin{center}
\epsfxsize=2.5in
\hspace*{0in}
\epsffile{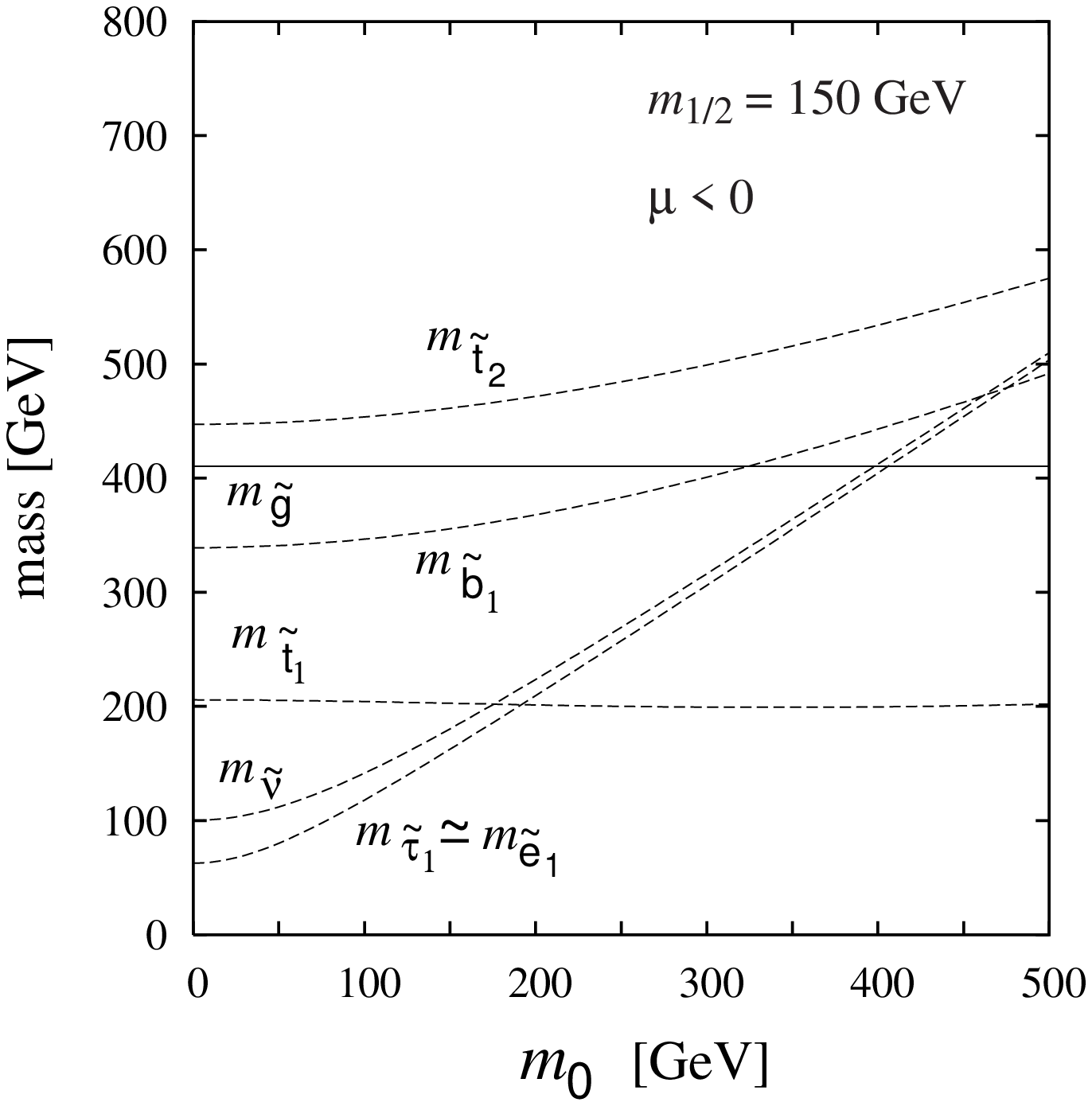}

\parbox{3.25in}{\small Fig.~5. The squark and slepton masses are plotted
versus $m_0^{}$ for $m_{1/2}=150$ GeV for a low value of $\tan \beta$ and
$\mu < 0$.}
\end{center}

The MSSM has an R-parity symmetry so the
lightest supersymmetric particle (LSP) is stable. Usually the LSP is the
lightest neutralino, but for small values of $m_0$ the supersymmetric partner
of the tau lepton can be lighter. For the lightest particle to be neutral, as
required, there is an upper bound on the value of $m_{1/2}$ for small
$m_0$. In particular such an upper bound exists for no-scale models ($m_0=0$),
and is more stringent for $\mu >0$ due to the mixing between the left- and
right-handed $\tilde{\tau }$. The phenomenological
issues of the Yukawa unified no-scale model have been
examined in Ref.~\cite{gunionpois}.

 The LSP can naturally account for the dark matter of the
Universe\cite{dm,RoRo}.
Large values of $\mu $ result in the lightest neutralino being
predominantly gaugino. This leads to a reduced rate of
annihilation of neutralinos and can provide too much relic abundance and
overclose the Universe. However the $s$-channel $h$ pole can enhance the
annihilation rate and rescue the dark matter explanation\cite{arnath-dm}.

\section{Proton Decay}

One of the major additions to particle physics from the concept of grand
unified theories is that the proton might be unstable. Stringent experimental
limits on proton decay rule out many models, including
the nonsupersymmetric version of SU(5). At first sight it might
appear that the supersymmetric versions of SU(5) are safe, since the
GUT scale is considerably higher and therefore proton decay
occurring through the dimension six operators is much suppressed.
However, dangerous dimension five operators are introduced in the
supersymmetric versions of grand unified models. In the minimal supersymmetric
SU(5) GUT, one must have very heavy sleptons to avoid the proton decay
bound\cite{an,hmy}. However, other models can greatly suppress or eliminate
entirely proton decay.

\section{Possibilities for Experimental Searches}

There are many interesting signals for supersymmetry at present and future
colliders. The missing $p_T$ signal at the Tevatron or at the LHC is a
classic experimental signature of supersymmetry. If the charginos and
the neutralinos are sufficiently light, then trilepton signals are
possible\cite{trilepton}:
\begin{eqnarray}
&&W^{\pm *}\to \chi^{\pm }_1\chi _2^0\to \ell^{\pm }\ell ^+\ell ^-\chi^0_1
\chi^0_1\;.
\end{eqnarray}
Gluinos will be produced abundantly at hadron colliders, and decays can
produce like-sign dilepton signals\cite{dileptons}. On the other hand, for
some regions of parameter space the gluino may decay predominantly into
stop~\cite{bdknt}:
\begin{eqnarray}
&&\tilde{g}\to \tilde{t}t\;.
\end{eqnarray}
If the stop is lighter than the top then stoponium bound states can be
 formed  which subsequently
decay into photon pairs or Higgs bosons\cite{stoponium}.
A future
high energy $e^+e^-$ collider (NLC) would provide an opportunity to
produce and study the properties of sleptons, charginos, and supersymmetric
Higgs bosons\cite{nlc}.

\section{Implications for $b\to s\gamma $ decay}

The measured rate
for the inclusive decay $b\to s\gamma $ \cite{cleo}
\begin{eqnarray}
{\rm BR}(B\to s\gamma)=(2.32\pm 0.51\pm 0.29\pm 0.32)\times 10^{-4}
\end{eqnarray}
is close to the
SM prediction. The predicted rate in SUSY models for small $\tan\beta$ is
somewhat  larger than the
SM for $\mu >0$ and generally smaller than the SM rate for $\mu < 0$\cite{bsg}.
Figure 6 shows the general trend of contours for the inclusive rate for
$\mu < 0$.
Unfortunately the current theoretical uncertainty is at least $\pm
25\%$\cite{bmmp},
so until further theoretical progress is made, one cannot determine the sign of
$\mu $.
SUSY contributions to $B^0-\overline{B}^0$, $D^0-\overline{D}^0$ and
$K^0-\overline{K}^0$ could also be relevant to placing restrictions on
models\cite{mixing}. The implications of the $Z\to b\overline{b}$ measurements
at LEP on supersymmetric unification have recently been
investigated\cite{park}.

\begin{center}
\epsfxsize=2.5in
\hspace*{0in}
\epsffile{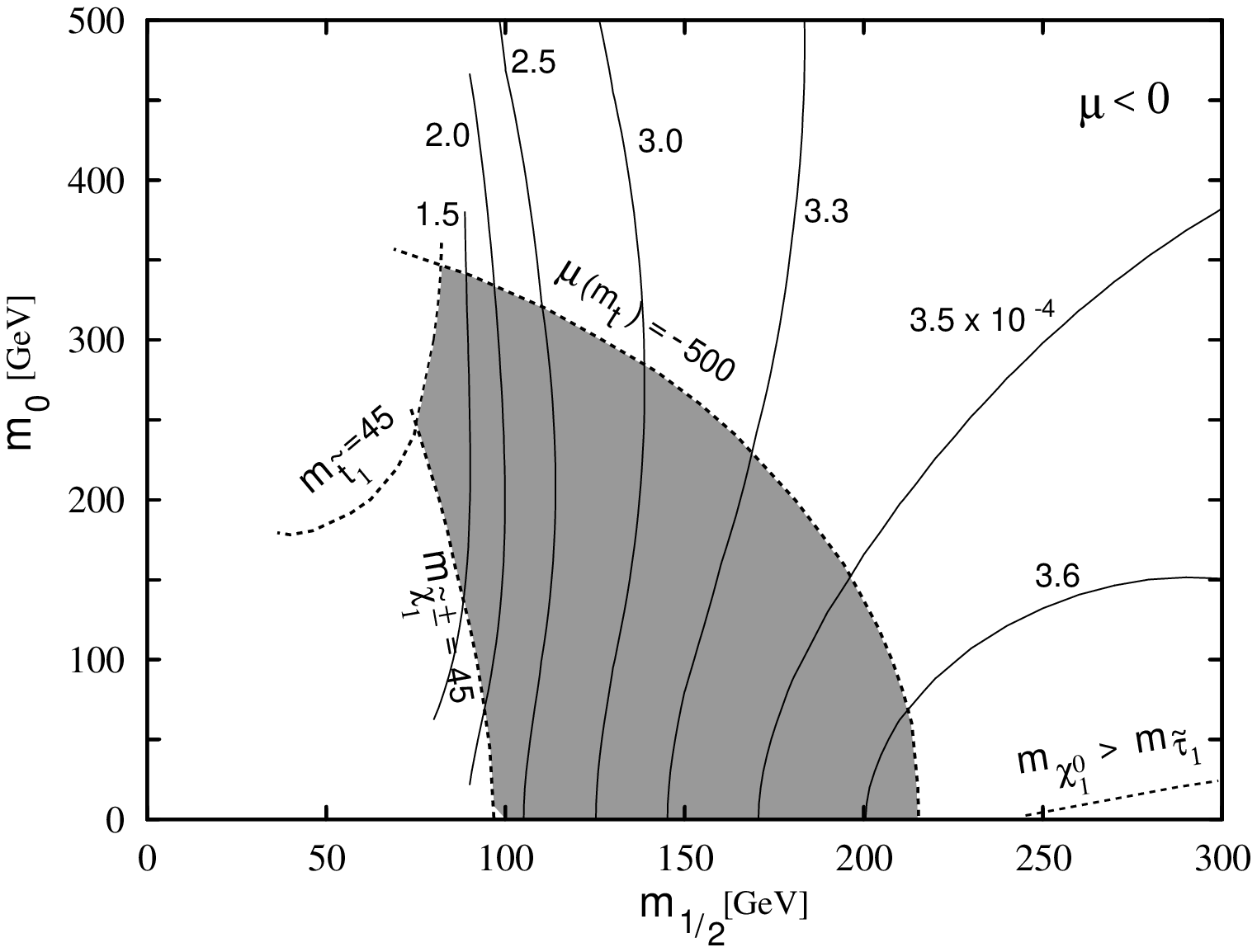}

\parbox{3.25in}{\small Fig.~6. Contour lines for the $b\to s\gamma $ inclusive
rate for $\mu < 0$ (from Ref.~\cite{bbopbsg}). }
\end{center}

\section{Conclusions}

According to all RGE sparticle mass spectrum analyses, two broad conclusions
about the implications of SUSY unification can be drawn:
\begin{itemize}

\item Interesting regions of the SUSY parameter space can be covered at the
Tevatron with the main injector, and possibly further improvements in the
luminosity or upgrades of the center-of-mass energy\cite{klmw}.

\item The LHC and the NLC are guaranteed to be SUSY factories if supersymmetry
exists; the task of determining how to pull the signals out of the backgrounds
is continuing\cite{factories}.

\end{itemize}

\section{Acknowledgements}

This research was supported
in part by the University of Wisconsin Research Committee with funds granted by
the Wisconsin Alumni Research Foundation, in part by the U.S.~Department of
Energy under contract nos.~DE-AC02-76ER00881 and DE-FG02-91ER40661,
and in part by the Texas National
Laboratory Research Commission under grant nos.~RGFY93-221.

\end{document}